\documentclass[aps,twocolumn,prl,showpacs]{revtex4}

\usepackage{amsmath,amssymb,amsthm,bbm,latexsym}    
\usepackage{amsthm} 
\usepackage{amsbsy} 
\usepackage[dvips]{graphicx}                
\usepackage{subfigure}                  
\usepackage{multirow}                   

\newcommand{\bra}[1]{\left\langle#1\right|} 
\newcommand{\ket}[1]{|#1\rangle} 
\newcommand{\brasmall}[1]{\langle#1|}   
\newcommand{\ketsmall}[1]{|#1\rangle}   

\newcommand{\bracket}[3]{\left\langle #1 \,\vrule\, #2 \,\vrule\, #3 \right\rangle}
\newcommand{\bracketsmall}[3]{\langle #1 \,|\, #2 \,|\, #3 \rangle}

\newcommand{\norm}[1]{\left|\left| #1\right|\right|}
\newcommand{\identity}{\mathbbm I}
\newcommand{\C}{\mathbbm C} 
\newcommand{\cd}[1]{c^\dag_{#1}}
\newcommand{\tr}{\mathrm{Tr}\,}

\newcommand{\mdash}{\textemdash}
\newcommand{\ie}{{\it i.e. }}
\newcommand{\eg}{{\it e.g. }}

\theoremstyle{plain}
\newtheorem{thm}{Theorem}
\newtheorem{cor}[thm]{Corollary}
\newtheorem{lem}[thm]{Lemma}
\theoremstyle{remark}

\theoremstyle{definition}

\newcommand{\ucl}{\affiliation{Department of Physics and Astronomy, University College London, Gower Street, London WC1E 6BT, United Kingdom}}

\begin{document}

\title{Multi-level, multi-party singlets as ground states\\and their role in entanglement distribution}
\author{Christopher Hadley}\ucl
\author{Sougato Bose}\ucl
\date{\today}

\begin{abstract}
We show that a singlet of many multi-level quantum systems arises naturally as the ground state of a physically-motivated Hamiltonian.  The Hamiltonian simply exchanges the states of nearest-neighbours in some network of qudits ($d$-level systems); the results are independent of the strength of the couplings or the network's topology.  We show that local measurements on some of these qudits project the unmeasured qudits onto a smaller singlet, regardless of the choice of measurement basis at each measurement.  It follows that the entanglement is highly {\it persistent}, and that through local measurements, a large amount of entanglement may be established between spatially-separated parties for subsequent use in distributed quantum computation.  
\end{abstract}
\maketitle

Entanglement between spatially-separated systems is a pivotal resource in quantum information theory, enabling distributed or networked quantum computation.  Thus there is an enormous interest in extracting this resource from the ground states of many-body systems \cite{article2007amico-fazio-osterloh-vedral}, in particular through measurements \cite{article2004verstraete-popp-cirac}. In such schemes, the measurement bases have to be carefully optimised. Could there be other systems whose ground states offer a more flexible method of entanglement extraction, and could the amount of entanglement exceed the currently-known limits?  Here we show that this is indeed the case for a particular system: a frustration-free {\it permutation} Hamiltonian \cite{article1975sutherland, article2005chen-wang, article2001shen, article2000li-gu-ying-eckern}, obtainable from the many-species Hubbard model (either fermionic or bosonic).  Such Hamiltonians have provoked much interest in many-body theory, since permutation Hamiltonians are a generalisation of the Heisenberg model \cite{article1975sutherland}, and the Hubbard model may be used to model phenomena ranging from superconductivity to ultracold atoms in optical lattices. We show that the ground state of this Hamiltonian is the $N$-partite singlet made of $N$-level systems, independently of both the spatial arrangement of the subsystems, and the strength of their couplings.  

The singlet is a highly-entangled, curious state, and has been shown to have wide-ranging applications in quantum information: in problems lacking classical solutions \cite{article2002cabello}, for multi-party remote state preparation \cite{article2003agrawal-parashar-pati}, and for encoding qubits in decoherence-free subspaces \cite{article2002cabello,article2002kok-nemoto-munro}.  In this Letter, we further consider some of the state's properties in the context of quantum information.  We show that if $N$ parties share an $N$-level singlet, and $M$ parties perform successive measurements (each in a random basis), the remaining parties share a singlet of $N-M$ systems, regardless of the choice of measurement bases (measurement in the same basis has been previously considered \cite{article2002cabello}). A direct consequence of this is that these states have the highest possible {\it persistency of entanglement} (the robustness of a multi-party entangled state to local measurements \cite{article2001briegel-raussendorf}).  This basis-independence is particularly interesting in the context of entanglement distribution: as already stated, entanglement can be established between spatially-separated parties (for example, between the extremal spins of a 1D spin chain) by performing local measurements on the other, intervening, systems. One can define the {\it localisable entanglement} as the maximum entanglement establishable in this way \cite{article2004verstraete-popp-cirac}. In general, maximising this quantity requires careful optimisation of the local measurement bases, and averaging over all outcomes.  However, for the singlet, even if one selected the basis {\it randomly} at each measurement, one can always establish a maximally-entangled state between the unmeasured parties.  This is a qualitative difference in mechanism of entanglement localisation to those systems studied previously \cite{article2004verstraete-popp-cirac}.  Moreover, another important difference emerges if one considers the entanglement establishable between two {\it subsystems} (\ie groups of constituent systems), rather than two individual constituents, of a system (\eg blocks of spins in a 1D chain, rather than individual spins).  In models studied so far, the entanglement establishable between two subsystems has no reason to differ from that of two individual parts; however, in our model two $n$-particle subsystems may share a maximally-entangled state of dimension $\binom{2n}{n}$ (equivalent to $\log_2 \binom{2n}{n}$ EPR singlets).  We also show that the {\it block entropy} of an $N$-singlet (the entanglement of a subsystem and the remainder of the system) is $\log_2 \binom{N}{L}$ where $L$ is the block size.  Finally, we discuss a potential physical realisation of a singlet in an optical lattice.  



{\it Qudit singlets:} A qudit is a generic $d$-level system, and qudit singlets $\ketsmall{S^{(d)}_N}$ are $N$-partite states with the property $U^{\otimes N}
\ketsmall{S^{(d)}_N}=\ketsmall{S^{(d)}_N}$ (up to a global phase), where $U$ is an arbitrary one-qudit unitary operation. For the case $d=N$, they may be written
\begin{align}
\ket{S^{(N)}_N ({\boldsymbol \alpha})} = \frac{1}{\sqrt{N!}}\sum_{\{n_l\}} \epsilon_{n_1,\cdots ,n_N} \ket{\alpha_{n_1},\cdots ,\alpha_{n_N}},\nonumber
\end{align}
where $\epsilon_{n_1,\cdots ,n_N}$ is the generalised Levi-Civita symbol, and the state is written in the basis $\{\ket{\alpha_i}\}_{i=1}^d$ at each qudit.  The sum is taken over all permutations of incides $\{n_1,\cdots ,n_N\}$, and crucially, is anti-symmetric with respect to the exchange (permutation) of any two qudits: \ie $P_{ij}\ket{S^{(N)}_N ({\boldsymbol \alpha})}=-\ket{S^{(N)}_N ({\boldsymbol \alpha})}$ for $i,j\in [1,N]$ where for given states $\ket{\psi}_i\in \mathcal{H}_i, \ket{\phi}_j\in \mathcal{H}_j$, the permutation operator $P_{ij}$ swaps the states according to $P_{ij}\ket{\psi}_i\ket{\phi}_j=\ket{\phi}_i\ket{\psi}_j$ and has eigenvalues $\pm 1$.

{\it Permutation Hamiltonian ground states:} The familiar isotropic Heisenberg Hamiltonian (${\boldsymbol \sigma}^i\cdot {\boldsymbol \sigma}^j$) acting on two qubits $i$, $j$ is equivalent (when the two-qubit identity operator is added) to the permutation operator $P_{ij}$.  The natural generalisation of this Hamiltonian for $d$-level systems is a sum of permutation operators $P_{ij}$ \cite{article1975sutherland}.  For a general network of qudits (some arrangement of qudits connected by permutation operators) one may use the language of {\it graph theory} and ascribe a finite graph $G:=\{V(G),E(G)\}$, where $V(G)$ denotes its set of vertices and $E(G)$ its set of edges\mdash if $(i,j)$ are adjacent vertices, $(i,j)\in E(G)$.  At each vertex we associate a $d$-level Hilbert space.  We avoid the term {\it lattice} at this stage, since this implies a regular arrangement of qudits; these results hold for more general networks, making graph theory the natural description.  The Hamiltonian is then written $H = \sum_{i,j\in E(G)} J_{ij}P_{ij}$; this may be obtained from the many-species Hubbard model (discussed below).  We now show that for these Hamiltonians, with $d=N$ and all couplings $J_{ij}>0$ (antiferromagnetic), the qudit singlet $\ketsmall{S^{(N)}_N}$ arises as the ground state.  We term this an `$N$-singlet' \footnote{The general case $d\ne N$ requires the use of the Bethe Ansatz \cite{article1975sutherland}.}.


\begin{lem}
The lowest possible energy state of a permutation Hamiltonian has energy equal to that of an eigenstate of all $\{P_{ij} | i,j\in E(G)\}$ with eigenvalue $-1$.
\begin{proof}
By definition, the ground state must minimise the energy $\bracket{\psi}{H}{\psi}$.  Now $\min_{\ket{\psi}\in (\C^d)^{\otimes N}} \bracket{\psi}{H}{\psi} \ge \sum_{i,j\in E(G)} \min_{\ket{\psi}\in (\C^d)^{\otimes N}} \bracket{\psi}{J_{ij}P_{ij}}{\psi}.$  The smallest eigenvalue of $P_{ij}$ is $-1$, thus $\min\bracket{\psi}{P_{ij}}{\psi}=-1$; it follows that $\min_{\ket{\psi}\in (\C^d)^{\otimes N}} \bracketsmall{\psi}{H}{\psi} \ge -\sum_{i,j}J_{ij}$.  Equality exists for an eigenstate of all terms in the Hamiltonian (\ie $\{P_{ij}|i,j\in E(G)\}$) and if this state exists, it is the ground state.
\end{proof}
\end{lem}

\begin{lem}\label{lemmaeigenstate}
If a state is an eigenstate of $\{ P_{ij} | i,j\in E(G) \}$, it is an eigenstate of all $\{ P_{ij} | i,j\in [1,N]\}$.
\begin{proof}
Consider first a linear chain with nearest-neighbour permutations.  Any other permutation may be written as a product of an odd number of nearest-neighbour permutations, (\eg $P_{13} = P_{23}P_{12}P_{23}$; generally $P_{1,k} = P_{1,k-1}P_{k-1,k}P_{1,k-1}$).  Thus an eigenstate of all nearest-neighbour permutations must also be an eigenstate of {\it all} possible permutations. This can be readily generalised to any connected graph, since any `path' through the graph is equivalent to a 1D chain of nearest-neigbour connections; thus any permutation in any graph can be written as a product of an odd number of nearest-neighbour permutations.\end{proof}
\end{lem}

\begin{thm}
The ground state of a permutation Hamiltonian on $N$ $N$-level systems is an $N$-singlet.\label{theorem-groundstate}
\begin{proof}
In the above lemmata we have shown that a state antisymmetric under all permutations is a valid ground state; since a qudit singlet for $d=N$ by definition satisfies this, it is a valid ground state.  It can easily be shown to be the {\it unique} ground state by assuming the existence of another distinct ground
state $\ket{\phi} \ne \ketsmall{S^{(N)}_N}$. Being a ground state, $\ket{\phi}$ must minimise the total energy, leading to $\bracket{\phi}{P_{ij}}{\phi}=-1$ for all $i,j$.  This implies $\ket{\phi} = \ketsmall{S^{(N)}_N}$. This is a contradiction and thus there cannot be another ground state.\end{proof}\end{thm}

Having shown how such singlets arise as ground states, we now go on to consider some of their properties and applications.

{\it Local measurements:} 
We show that when $N$ parties share such a state, and some of them perform measurements on their qudits in randomly chosen bases (varying from party to party), a qudit
singlet is still established between those unmeasured qudits, in a basis related to that used for the last measurement.  To demonstrate this, we will make use of the property
\cite{article2003agrawal-parashar-pati} $U\otimes\identity^{\otimes N-1} \ketsmall{S^{(N)}_N} = \identity\otimes U^{\dag\otimes N-1}\ketsmall{S^{(N)}_N}$ and introduce the notation $\ketsmall{S^{(N-1)}_{N-1} ({\boldsymbol\beta};\beta_l)}$ to denote an $(N-1)$-singlet written in the basis $\{\ket{\beta_i}\}_{i=1}^N$ at each qudit, with level $\ket{\beta_l}$ absent (\ie $\tr \rho_i\ketsmall{\beta_l}\brasmall{\beta_l} = 0$ for all $i$ where $\rho_i$ is the reduced density matrix of the $i$th qudit).
\begin{thm}
\label{theorem-measure}
When an $N$-singlet $\ketsmall{S^{(N)}_N ({\boldsymbol\alpha})}$ written in a basis $\{\ket{\alpha_i}\}$ is measured at one site using an arbitrary basis $\{\ket{\beta_i}\} := \{U\ket{\alpha_i}\}$, the state obtained is a product of a state $\ket{\beta_i}$ at the measured site and a smaller singlet $\ketsmall{S^{(N-1)}_{N-1} ({\boldsymbol\beta};\beta_i)}$ written in the basis $\{\ket{\beta_i}\}$.  
\begin{proof}
Consider the outcome when we perform a von Neumann measurement $\{\ket{\beta_i}\bra{\beta_i}\}$ at one qudit.  Without loss of generality, we take the above measurement on the qudit labelled $1$. Since $\ket{\beta_i}\bra{\beta_i} = U\ket{\alpha_i}\bra{\alpha_i}U^\dag$ we may write
\begin{align}
\ket{\beta_i}\bra{\beta_i}&\otimes\identity^{\otimes N-1}\ket{S^{(N)}_N ({\boldsymbol\alpha})}\nonumber\\
&= U\ket{\alpha_i}\bra{\alpha_i}U^\dag\otimes\identity^{\otimes N-1}\ket{S_N^{(N)} ({\boldsymbol\alpha})}\nonumber\\
&= U\ket{\alpha_i}\bra{\alpha_i}\identity\otimes U^{\otimes N-1}\ket{S_N^{(N)} ({\boldsymbol\alpha})}.\label{thmtransform}
\end{align}
To proceed, we write the $N$-singlet as
\begin{align}
\ket{S^{(N)}_N ({\boldsymbol\alpha})} = \frac{1}{\sqrt{N}}\sum_{i=1}^{N} (-)^{i+1}\ket{\alpha_i}_1\ket{S^{(N-1)}_{N-1}({\boldsymbol \alpha};\alpha_i)}_{2,\cdots ,N},\nonumber
\end{align}
and it thus follows that
\begin{align}
\identity&\otimes U^{\otimes N-1}\ket{S^{(N)}_N({\boldsymbol \alpha})}_{1,\cdots ,N}\nonumber\\
&= \frac{1}{\sqrt{N}}\sum_{i=1}^N (-)^{i+1} \ket{\alpha_i}_1\otimes U^{\otimes N-1}\ket{S^{(N-1)}_{N-1}(\boldsymbol{\alpha};\alpha_i)}_{2,\cdots N}.\nonumber
\end{align}
Whilst $\ketsmall{S^{(N-1)}_{N-1}({\boldsymbol\alpha};\alpha_i)}$ is a singlet written in the $\{\ket{\alpha_l}\}_{l\ne i}$ basis at each qudit, one can see that the term $U^{\otimes N-1}\ketsmall{S^{(N-1)}_{N-1}(\boldsymbol{\alpha};\alpha_i)}$ is a singlet written in the $\{\ket{\beta_l}\}_{l\ne i}$ basis \footnote{These are {\it not} the same state, since the bases are not complete in the $N$-dimensional space.}.  Thus if the measurement outcome is $\ket{\beta_i}$, the overall state is projected to
\begin{align}
\ket{\beta_i}\bra{\beta_i}&\otimes\identity^{\otimes N-1}\ket{S^{(N)}_N ({\boldsymbol\alpha})}_{1,\cdots,N} / \norm{...}\nonumber\\
&= (-)^{i+1}\ket{\beta_i}_1\otimes\ket{S^{(N-1)}_{N-1}(\boldsymbol{\beta};\beta_i)}_{2,\cdots ,N}.\nonumber
\end{align}
This completes the proof.\end{proof}\end{thm}

The significance of the above theorem is revealed when one considers
successive measurements at different qudits. Indeed, by iterating
the above proof, it becomes apparent that by measuring in a
different basis at each qudit, the remaining, unmeasured qudits will
be projected to a singlet in a basis related to that used for the
final measurement.

\begin{cor}
If $M$ parties perform successive measurements in arbitrary bases
(where the $m$th party uses the basis $B_m =
\{\ketsmall{\alpha_i^{(m)}}\}_{i\ne 1,\cdots ,m-1} = \{
\prod_{l=1}^m U^{(l)}\ketsmall{\alpha^{(0)}_i} \}$), the remaining
parties share an $(N-M)$-singlet in the $B_M$ basis at each qudit,
with the restriction that at each measurement, the basis
transformation operates on a space whose dimension is one less than
the previous transformation; \ie at the $l$th measurement the basis
may be transformed by any $U^{(l)}$ acting on the subspace
$\C^{N-l}$.
\begin{proof}
Consider measuring in the basis $\{\ketsmall{\alpha^{(2)}_i}\}$ on the state resulting from the previous measurement, namely, $\ketsmall{\alpha^{(1)}_i}_1\ketsmall{S^{(N-1)}_{N-1}(\boldsymbol{\alpha^{(1)}};\alpha^{(1)}_i)}_{2,\cdots ,N}$.  Repeating Theorem \ref{theorem-measure} on the smaller singlet, one can see that the final state is $\ket{\alpha^{(1)}_i}_1\ket{\alpha^{(2)}_j}_2\ket{S^{(N-2)}_{N-2}(\boldsymbol{\alpha^{(2)}};\alpha^{(2)}_i,\alpha^{(2)}_j)}_{3,\cdots ,N}$.  In general, if $M$ measurements are taken, we have (upto a global phase) $
\ket{\alpha^{(1)}_{n_1}}_1\cdots\ket{\alpha^{(M)}_{n_M}}_M\ket{S^{(N-M)}_{N-M} ( {\boldsymbol\alpha^{(M)}_{n_M}} ; {\boldsymbol {\mathrm n}^{(M)}}     )   }_{M+1,\cdots ,N}$ where ${\boldsymbol {\mathrm n}} = (n_1,\cdots ,n_M)$ is a vector, the elements of which are the indices of the
vectors of the basis $B_M$ excluded from the $(N-M)$-singlet.
\end{proof}
\end{cor}

It must be noted that in the proof\mdash see equation (\ref{thmtransform})\mdash we make use of the property $U^{\otimes N}\ketsmall{S^{(N)}_N} =\ketsmall{S^{(N)}_N} $. This holds only when the $U^{\otimes N}$ operates on the space occupied by the singlet $\ketsmall{S^{(N)}_N}$; if one were to put this state within a larger space, a separable unitary operation on the whole $n$-partite space would not (in general) give this invariance.  Thus in order to iterate the proof, we make the restriction that at each successive measurement the dimension of the unitary transformation decreases by one.  So, at the $l$th measurement, one would be able to transform the basis by a unitary  $U^{(l)}$, such that there is a submatrix operating on $N-l$ levels (which levels are operated on depends on the previous outcomes), and `1' on all diagonal elements corresponding to the remaining levels.
In fact, this restriction may be lifted slightly. It is well known
\cite{article1994reck-zeilinger-bernstein-bertani} that any $d\times
d$ unitary matrix can be written as a product of two-level
unitaries: $U_d = V_1 \cdots V_k$, where $k\le d(d-1)/2$.  This then gives
\begin{align}
U_d\otimes U_d = (V_1\otimes V_1)\cdots (V_k\otimes V_k),\label{factoroperator}
\end{align}
allowing one to use at measurement the subset of $d\times d$ unitaries that factorise such that the two-level matrices $\{V_i\}$ either act within the subspace supporting the singlet, or its complement (\ie the matrix element linking the singlet subspace to the rest of the space is zero). This is because those factors operating on the complement act as the identity on the singlet, and those operating on the singlet subspace have the rotational invariance property.

But what would happen if one measured in an arbitrary basis?  To answer this, consider the effect of operating on a 2-singlet with a more general unitary operator, written in the form (\ref{factoroperator}).  Those factors $V_i\otimes V_i$ that do not operate on any part of the singlet subspace can be removed successively from the right of the operator (\ref{factoroperator}), until the right-most factor links one of the levels of the singlet with a different level.  Suppose the $i$th factor in (\ref{factoroperator}) operates on levels $j,k$ in a basis $\{\ket{l}\}_{l=1}^d$, and the whole operator $U_d\otimes U_d$ is applied to a singlet of levels $j,m$.  Then $(V_1 \otimes V_1)\cdots (V_{i}\otimes V_{i})(\ket{jm}-\ket{mj}) = (V_1\otimes V_1)\cdots (V_{i-1}\otimes V_{i-1})   (                  (V_i\ket{j})\ket{m}-\ket{m}(V_i\ket{j})  )$.  This remains a singlet, albeit in a different subspace.  In general, when an $n$-singlet is operated on by a separable unitary $U_d^{\otimes n}$ (where $d>n$), the state remains an $n$-singlet, but (in general) lies within a different subspace of the $n^n$-dimensional $n$-partite Hilbert space acted on by $U_d^{\otimes n}$.  Thus we conclude that for any measurement, we still obtain a singlet, but for those measurement bases not satisfying the constraints given above, the singlet moves into a different subspace.

{\it Entanglement properties:} 
Though there is a plethora of entanglement measures for many-body systems, perhaps the most physically-relevant is the amount of entanglement one can establish between spatially-separated subsystems for subsequent use in networked quantum computation.  Indeed, it was with this motivation that {\it localisable entanglement} was introduced \cite{article2004verstraete-popp-cirac} and a maximally-entangled state of a $2\times 2$ system has been shown to be
establishable this way. However, any realistic distributed quantum computation scheme would require a larger amount of more entanglement.  Using the above statements, we can now consider the role of $N$-singlets in entanglement distribution.  Suppose that Alice and Bob each hold $n$ qudits of an $N$-singlet.  By measuring the remaining qudits, they can establish a $2n$-singlet between them.  This is equivalent to sharing a maximally-entangled state of dimension $D\times D$ (where $D:=\binom{2n}{n}$) or $\log_2 D$ EPR singlets, since the $2n$-singlet can be written in a Schmidt decomposition of $\binom{2n}{n}$ terms of equal amplitude (see Fig. \ref{LE}), and upto local unitaries this is equivalent to $\sum_{i=1}^D\ket{i}\ket{i}/\sqrt{D}$. Thus the localisable entanglement in our model (when the notion is generalised to two subsystems) can be much larger in comparison to the models studied so far.  Furthermore, by taking a Schmidt decomposition across an arbitrary partitioning of qudits, one can see that the {\it block entropy} of this state is $\log_2 \binom{N}{L}$.  This is in contrast to the usual behaviour for gapped systems, where the entropy is usually proportional the block's `area', and to the case of gapless spin chains in 1D where it is proportional to $\log_2 L$.  We finally state without proof that the  $N$-singlet is genuinely $N$-partite entangled (\ie the state is non-separable with respect to all possible partitionings of qudits, and it is not producible by $k$-partite entanglement, for all $k<N$).
\begin{small}\begin{figure}
 \begin{center}\begin{tabular}{c}
   \includegraphics[width=6.5cm,angle=0]{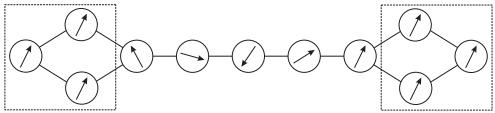}
  \end{tabular}\end{center}
 \caption{Two subsystem localisable entanglement.  Here, Alice and Bob have access to the boxed qudits.  By performing arbitrary measurements on the other qudits (in random directions), they can establish a 6-singlet between them.}
 \label{LE}
\end{figure}\end{small}

{\it Persistency of entanglement:}  An important consideration from a practical point of view is the ease of destroying the entanglement present.  This can be quantified by the {\it persistency of entanglement}, the minimum number of local von Neumann measurements that will disentangle the state \cite{article2001briegel-raussendorf}.  This may be used to model an environment that interacts with a system through random, local measurements.  Since the establishment of successively smaller singlets is {\it independent} of both the basis choice and outcome, it is easy to see that $N-1$ measurements always need to be performed to completely disentangle the state (there is no way to optimise the basis to reduce this number).  Thus these states have the highest possible persistency of an $N$-partite state (for comparison, the multi-qubit cluster, GHZ and $W$ states have persistencies $\lfloor N/2\rfloor$, $1$ and $N-1$, respectively).

{\it Physical preparation of states in 1D:}
The permutation Hamiltonian may be obtained as an effective Hamiltonian in a certain regime of the $d$-species Hubbard model:
\begin{align}
H = - \sum_{i,j}\sum_{\sigma = 1}^d t^\sigma_{ij}\left(
c^\dag_{\sigma i} c_{\sigma j} + c^\dag_{\sigma j} c_{\sigma i}
\right) + \sum_i \sum_{\sigma\ne\sigma '} U_{\sigma,\sigma'} n_{\sigma i} n_{\sigma
'i},\nonumber
\end{align}
where $i$, $j$ are summed over some network, $U_{\sigma,\sigma'}$ is the interaction between species $\sigma$ and $\sigma'$, $t^\sigma_{ij}$ the hopping integral between lattice sites $i$ and $j$ for species $\sigma$, and $c^\dag_{\sigma i}$ is the creation operator for $\sigma$ at lattice site $i$ (either fermionic or bosonic).  {\it Fermionic:} In the regime of strong, isotropic couplings ($U\gg t_{ij}$), with $1/d$-filling regime, this (following a similar perturbation calculation to Fradkin \cite{book1991fradkin} and Emery \cite{article1976emery} for the two-level case) becomes $H' = \sum_{\alpha ,\beta =1}^d \sum_{i,j} J_{ij} \cd{\alpha i}c_{\beta i}\cd{\beta j}c_{\alpha j}$, exactly the permutation Hamiltonian required with $J_{ij}=4t_{ij}^2/U$ \footnote{See EPAPS Document No. E-PLRAAN-77-R17804 for further details on the derivation of the effective permutation Hamiltonian from the Hubbard model, and the genuine $N$-partite nature of the entanglement.  For more information on EPAPS, see http://www.aip.org/pubservs/epaps.html.}.  {\it Bosonic:} the permutation Hamiltonian is realisable in the $1/d$-filling regime with careful tuning of the species-dependent tunneling integrals and the addition of a hard-core interaction \cite{article2003kuklov-svistunov}.  

This is one of the most widely-studied condensed matter models, and has been shown to be realisable with ultracold atoms in optical lattices.  Two-level atoms at $1/2$-filling realise the Heisenberg model \cite{article2003kuklov-svistunov}, and replacing these with atoms with $d$ degenerate levels (\eg $d$ hyperfine levels \cite{article1995parkins-marte-zoller-carnal-kimble} where $d=10$ is routine with $^{40}$K \cite{article2005koehl-moritz-stoeferle-guenter-esslinger} and much higher $d$ is possible with Er \cite{article2006mcclelland-hanssen}) realises the permutation Hamiltonian.  To prepare the singlet, a $d$-site optical lattice is first loaded with $d$-level atoms in the translationally-invariant state $\ket{1}^{\otimes d}$ (with all $t_{ij}\sim 0$).  An inhomogenous $B$-field coupled to the hyperfine levels is applied to select this as the unique ground state within the $1/d$-filling regime and to allow parity to be broken in the subsequent evolution.  The ratios $t_{ij}/U$ are then increased adiabatically (the standard method for ground state preparation in optical lattices \cite{article2002greiner-mandel-esslinger-haensch-bloch}) to reach finite, but small $t_{ij}\ll U$ (still in the Mott insulator regime); similarly the $B$-field is slowly tuned to zero.  The system's state evolves to the final ground state: the singlet.  Other potential physical implementations include spin ladders and tubes \cite{article1999batchelor-maslen}, arrays of quantum dots with both spin and orbital levels \cite{article2000li-gu-ying-eckern}.  To date, little progress has been made in realisation, although entangled photons \cite{article2001lamaslinares-howell-bouwmeester} and cavity QED \cite{article2005jin-li-feng-zheng, article2007lin-ye-chen-du-lin} have been considered.


{\it Acknowledgements:} C.H. acknowledges financial support from the
EPSRC through grant EP/P500559/1. We thank J. I. Cirac, A. J. Fisher, V. E. Korepin and A. Serafini for useful discussions.




\end{document}